\documentclass[apj]{emulateapj}
\usepackage{apjfonts}

\gdef\kms{km\,s$^{-1}$}
\gdef\msun{$M_{\odot}$}
\lefthead{van Dokkum et al.}
\righthead{Sizes of Quiescent Galaxies at $z\sim 2.3$}
\slugcomment{Accepted for publication in ApJ Letters}
\begin{document}

\title{Confirmation
of the remarkable compactness of massive quiescent galaxies
at $z\sim 2.3$: early-type galaxies did not
form in a simple monolithic collapse
\altaffilmark{1,2}}

\author{Pieter~G.~van Dokkum\altaffilmark{3}, Marijn Franx\altaffilmark{4},
Mariska Kriek\altaffilmark{5}, 
Bradford Holden\altaffilmark{6}, Garth D.\ Illingworth\altaffilmark{6},
Daniel Magee\altaffilmark{6}, Rychard Bouwens\altaffilmark{6},
Danilo Marchesini\altaffilmark{3}, Ryan Quadri\altaffilmark{4},
Greg Rudnick\altaffilmark{7}, Edward N.\ Taylor\altaffilmark{4},
and Sune Toft\altaffilmark{8}}

\altaffiltext{1}
{Based on observations with the NASA/ESA {\em Hubble Space
Telescope}, obtained at the Space Telescope Science Institute, which
is operated by AURA, Inc., under NASA contract NAS 5--26555.}
\altaffiltext{2}
{Based on observations obtained with the W.\ M.\ Keck Observatory.
The Observatory was made possible by the generous financial support
of the W.\ M.\ Keck Foundation.}
\altaffiltext{3}
{Department of Astronomy, Yale University, New Haven, CT 06520-8101}
\altaffiltext{4}
{Sterrewacht Leiden, Leiden University, NL-2300 RA Leiden, Netherlands}
\altaffiltext{5}
{Department of Astrophysical Sciences, Princeton University, Princeton, NJ
08544}
\altaffiltext{6}
{UCO/Lick Observatory, University of California, Santa Cruz, CA 95064}
\altaffiltext{7}
{National Optical Astronomical Observatory, Tucson, AZ 85719}
\altaffiltext{8}
{European Southern Observatory, D-85748 Garching, Germany}

\begin{abstract}

Using deep near-infrared spectroscopy {Kriek} {et~al.} (2006) found that
$\sim 45$\,\%
of massive galaxies at $z\sim 2.3$ have evolved stellar
populations and little or no ongoing star formation.
Here we determine the sizes of these quiescent galaxies using
deep, high-resolution images
obtained with HST/NIC2 and laser guide
star-assisted Keck/AO.
Considering that their median stellar mass
is $1.7 \times 10^{11}$\,\msun\ the galaxies
are remarkably small, with a median effective radius
$r_e=0.9$\,kpc. Galaxies of
similar mass in the nearby Universe have sizes of
$\approx 5$\,kpc and average stellar densities which are
two orders of magnitude lower than the $z\sim 2.3$ galaxies.
These results extend earlier work at $z\sim 1.5$ and
confirm previous studies at $z>2$ which lacked
spectroscopic redshifts and imaging of sufficient resolution
to resolve the galaxies.
Our findings demonstrate that fully assembled early-type galaxies
make up at most $\sim 10$\,\%
of the population of $K$-selected quiescent
galaxies at $z\sim 2.3$,
effectively ruling out simple monolithic
models for their formation. The galaxies must
evolve significantly after $z\sim 2.3$, through dry mergers or other
processes, consistent with predictions from hierarchical models.

\end{abstract}


\section{Introduction}

The sizes of massive galaxies at high redshift
provide strong constraints on galaxy formation models. 
Models starting from $\Lambda$CDM initial conditions
have predicted that the progenitors of today's early-type galaxies
should be smaller by a factor of a few at redshifts of 2--3
(e.g., {Loeb} \& {Peebles} 2003; {Robertson} {et~al.} 2006; {Khochfar} \& {Silk} 2006; {Naab} {et~al.} 2007). This can be contrasted to
expectations from simple
``monolithic'' models (e.g., {Eggen}, {Lynden-Bell}, \&  {Sandage} 1962),
in which galaxies undergo little or no structural evolution since their
initial assembly at very high redshift.

Several recent studies have found evidence that massive galaxies
at high redshift are indeed quite compact, particularly
those with the lowest estimated star formation rates
({Daddi} {et~al.} 2005; {Trujillo} {et~al.} 2006, 2007; {Zirm} {et~al.} 2007; {Toft} {et~al.} 2007; Longhetti et al.\ 2007; {Cimatti} {et~al.} 2008).
The size evolution appears to be quite dramatic: {Toft} {et~al.} (2007)
find that the quiescent $z\sim 2.5$
galaxies in their sample have a factor of
30--40 higher surface density than red galaxies of the
same mass in the Sloan Digital Sky Survey (SDSS).

These results are still somewhat tentative, due to various systematic
uncertainties. First, the galaxies were selected in relatively
small fields, and therefore
typically do not span the mass range of today's giant
elliptical galaxies ($M \gtrsim 10^{11}$\,\msun). Second,
the imaging that has been used so far has either sampled the
rest-frame ultraviolet (e.g., {Cimatti} {et~al.} 2008), or has not
been of sufficient resolution to measure accurate rest-frame optical
sizes of very compact objects.
Third, some of the galaxies may have
a central star burst or active nucleus
which could skew their size measurements. Fourth, the
stellar ages and masses of the galaxies have large uncertainties,
as their broad-band photometry can typically be fitted by
a wide range of models (see, e.g., {Papovich}, {Dickinson}, \&  {Ferguson} 2001).
Lastly, and perhaps most importantly, studies at $z>2$ are typically
based on photometric redshifts, which are
poorly calibrated for faint, red galaxies. 

Recently {Kriek} {et~al.} (2006) used extensive near-infrared
spectroscopy to
secure the first sample of {\em spectroscopically confirmed}
massive ($M\gtrsim 10^{11}$\,\msun)
galaxies with evolved stellar populations at $z>2$. 
In this Letter, we present high quality {\it HST} NIC2 and Keck images
of these galaxies to
measure accurate rest-frame optical sizes.
The nine objects discussed here are the oldest objects in a complete
sample of $K$-selected massive galaxies at $z\sim 2.3$.
If early-type galaxies were already fully formed at this redshift
we should see no structural
differences between these galaxies and galaxies of similar mass at $z=0$.

\begin{figure*}[t]
\epsfxsize=16cm
\epsffile[4 136 528 626]{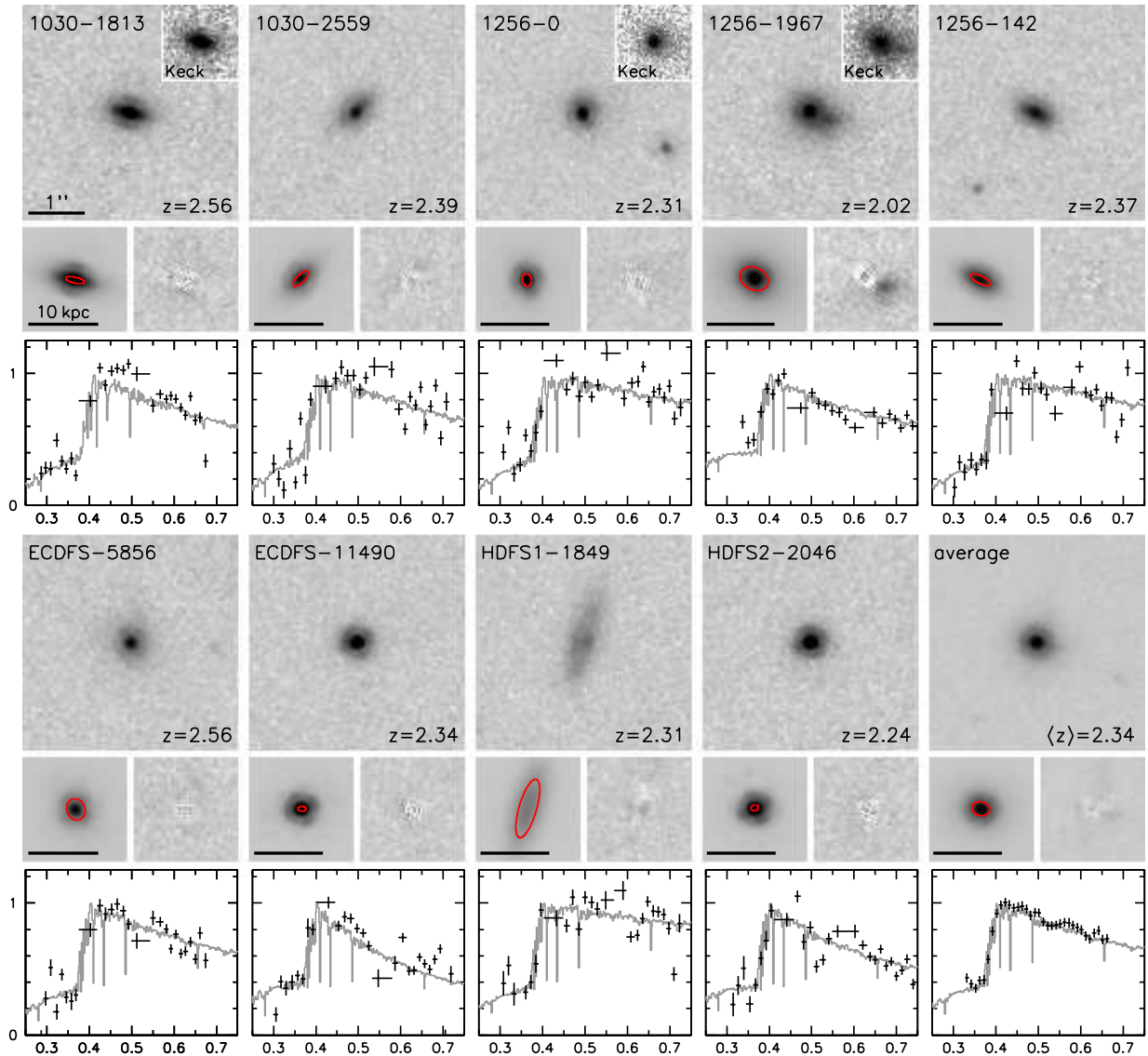}
\caption{\small
{\em HST} NIC2 images of the nine spectroscopically-confirmed quiescent $z>2$
galaxies from Kriek et al.\ (2006). Each panel spans $3\farcs 8
\times 3\farcs 8$; North is up and East is to the left.
The small panels below each galaxy
show the best-fitting Sersic model (convolved with the PSF) and
the residual after subtraction of the best-fitting model. The red
ellipses are constructed from the best-fitting effective radii,
axis ratios, and position angles. Note that the ellipses are
significantly smaller than 10\,kpc, which is the effective diameter of typical
massive elliptical galaxies in the nearby Universe. Gemini GNIRS
spectra from Kriek et al.\ (2006) are also shown. Insets show
Keck LGS/AO images of three galaxies.
\label{stamps.plot}}
\end{figure*}

\section{Sample, Observations, and Reduction}

Kriek et al.\ (2006, 2008a) present a deep near-IR spectroscopic
survey of $K$-selected galaxies at $2<z<2.7$ with
the Gemini Near-Infrared Spectrograph. The galaxies were
originally identified in the deep, wide MUSYC near-IR
imaging survey (Quadri et al.\ 2007).
One of the key findings of Kriek et al.\
is that nearly half of the galaxies
show no detected H$\alpha$ emission
and have spectra dominated by strong Balmer/4000\,\AA\
breaks. Nine of these quiescent objects were presented in
Kriek et al.\ (2006).
These nine galaxies have a median redshift
of 2.34, stellar masses $\gtrsim 10^{11}$\,\msun,
and ages of 0.5 -- 1 Gyr. They define a
tight red sequence in rest-frame $U-B$ (Kriek et al.\ 2008b).

All nine quiescent galaxies were observed in the F160W filter
using the NIC2 camera on {\em HST},
from June 2006 -- June 2007.
Two orbits were used for each of the brightest two
galaxies and three orbits for each of
the remaining seven. Each orbit was split in two (dithered)
exposures.
The reduction followed the procedures outlined in {Bouwens} \& {Illingworth} (2006),
and R.\ Bouwens et al., in preparation.
Before combining, the individual exposures
were drizzled to a new grid with $0\farcs 0378$ pixels
to ensure that the PSF is well sampled.
Images of the nine galaxies are shown in Fig.\ \ref{stamps.plot}.

Three of the galaxies (1030-1813, 1256-0, and 1256-1967)
were also observed
with Keck, using laser guide star assisted adaptive optics to correct
for the atmosphere. The data
were obtained on 2007 May 14 and
2008 January 13 using the NIRC2 wide field camera, which gives a
pixel size of $0\farcs 04$. The reduction
followed standard procedures for near-IR imaging data.
The Keck images are shown as insets in Fig.\ 1. They show
the same qualitative features as the NICMOS data.

\begin{figure*}[t]
\epsfxsize=16cm
\epsffile[-40 0 856 430]{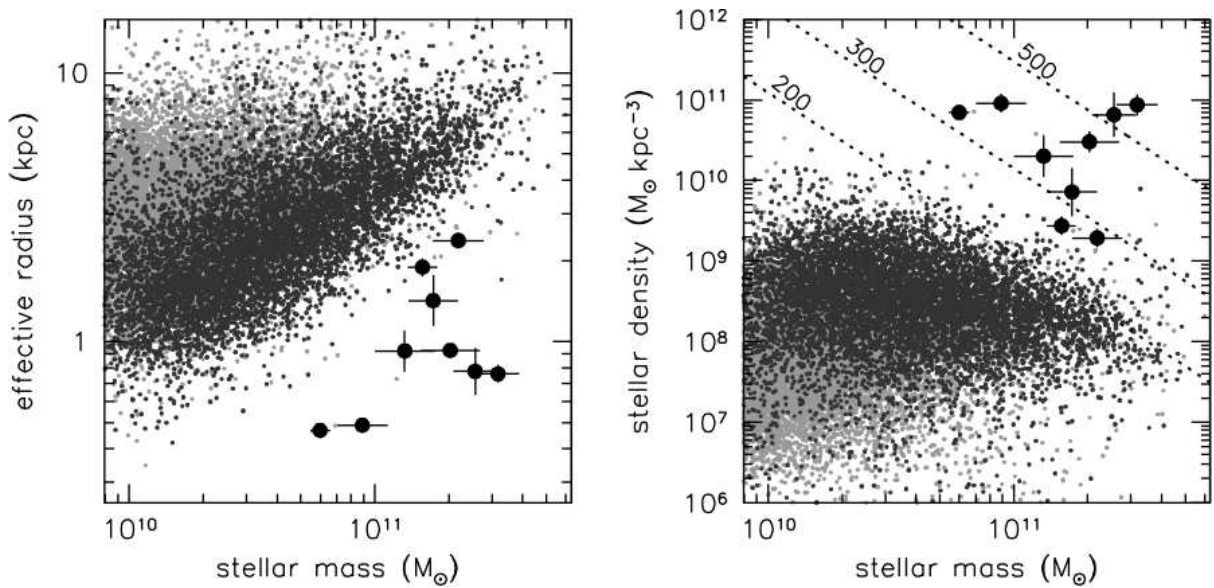}
\caption{\small
Relations between size and (total) stellar mass {\em (left panel)} and
between the average stellar density inside the effective radius
and stellar mass {\em (right panel)}. Large symbols
with errorbars are the quiescent $z\sim 2.3$ galaxies. Small symbols are
SDSS galaxies, with galaxies that are
not on the red sequence in light grey. Broken lines indicate the
expected location of galaxies with stellar velocity dispersions
of 200, 300, and 500\,\kms.
The high redshift galaxies are much smaller and denser than
SDSS galaxies of the same stellar mass.
\label{relations.plot}}
\end{figure*}

\section{Fitting}

Each galaxy was fitted with a {Sersic} (1968) radial surface brightness
profile, using the 2D fitting code
{\sc galfit} ({Peng} {et~al.} 2002). The Sersic $n$ parameter allows for
a large range of profile shapes, and provides a crude estimate
of the bulge-to-disk ratio.
For each galaxy a synthetic NIC2 PSF was created by generating subsampled
PSFs with Tiny Tim 6.3 ({Krist} 1995), shifting them
to replicate the location of the galaxy on the
individual exposures, binning these to the native NIC2
resolution, and finally drizzling these ``observations''
to the grid of the galaxy images.
The resulting fit parameters are listed in Table 1;
ellipses corresponding to the best-fit
parameters are indicated in red in Fig.\ \ref{stamps.plot}.
The (circularized) effective radii were transformed to kpc using
$H_0=70$\,\kms\,Mpc$^{-1}$, $\Omega_m=0.3$, and
$\Omega_{\Lambda}=0.7$.

Uncertainties in the structural parameters of faint galaxies
are difficult to estimate, as they are usually
dominated by systematic effects. For each galaxy, we added
the residual image of each of the other galaxies (excluding
1256-1967) in turn,
repeated the fit, and determined the rms of the
seven values obtained from these fits. The uncertainties
listed in Table 1 are $2\times$ these rms values, to account
for additional systematic uncertainties. These were assessed
by changing the size of the fitting region, scrambling the
subpixel positions of the galaxies, and changing the
drizzle grid.

\begin{small}
\begin{center}
{ {\sc TABLE 1} \\
\sc Structural Parameters} \\
\vspace{0.1cm}
\begin{tabular}{lccccccc}
\hline
\hline
ID & $z$ & $r_e^{(1)}$ & $\pm$ & $n$ & $\pm$ & $b/a$ & $\pm$ \\
\hline
1030-1813 & 2.56 & 0.76 & 0.06 & 1.9 & 0.5 & 0.30 & 0.03 \\ 
1030-2559 & 2.39 & 0.92 & 0.18 & 2.3 & 0.6 & 0.39 & 0.04 \\ 
1256-0 & 2.31 & 0.78 & 0.17 & 3.2 & 0.9 & 0.71 & 0.10 \\ 
1256-1967 & 2.02 & 1.89 & 0.15 & 3.4 & 0.1 & 0.75 & 0.07 \\ 
1256-142 & 2.37 & 0.93 & 0.04 & 0.9 & 0.3 & 0.35 & 0.04 \\ 
ECDFS-5856 & 2.56 & 1.42 & 0.35 & 4.5 & 0.4 & 0.83 & 0.07 \\ 
ECDFS-11490 & 2.34 & 0.47 & 0.03 & 2.8 & 0.8 & 0.63 & 0.07 \\ 
HDFS1-1849 & 2.31 & 2.38 & 0.11 & 0.5 & 0.2 & 0.29 & 0.02 \\ 
HDFS2-2046 & 2.24 & 0.49 & 0.02 & 2.3 & 0.8 & 0.76 & 0.08 \\
\hline
\end{tabular}
\end{center}
\vspace{-0.2cm}
\footnotesize{
(1) Circularized effective radius in kpc.}\\
\end{small}

The Keck images offer an independent test of the reliability
of the fit parameters. Fitting
the Keck images with a range of stellar PSFs
(including stars in the field-of-view)
gives results that are consistent with
the NIC2 fits within the listed uncertainties. As an example,
for 1030-1813 we find $r_e=0.73$\,kpc, $n=1.6$, and $b/a=0.32$
from the Keck image. In the following, we will use the values
derived from the higher S/N NIC2 images; our conclusions do
not change if we were to use the Keck results for
1030-1813, 1256-0, and 1256-1967.

\section{Sizes and Densities}

The most remarkable aspect of the $z\sim 2.3$ galaxies is their
compactness. The circularized effective radii range from 0.5 -- 2.4
kpc, and the median is 0.9\,kpc. To put this in context, this is
smaller than many bulges of spiral galaxies (including the bulges
of the Milky Way and M31, which have $r_e \approx 2.5$\,kpc;
{van den Bergh} 1999). In the left panel
of Fig.\ \ref{relations.plot} the sizes are compared to
those of SDSS galaxies.
The SDSS data were taken from the NYU Value Added Galaxy Catalog
({Blanton} {et~al.} 2005) in a narrow redshift range,
with various small corrections (see M.\ Franx et al., in preperation).
Dark grey points are galaxies on the red sequence, here
defined as $u-g = 0.1 \log (M) + (0.6 \pm 0.2)$.
Stellar masses for the $z\sim 2.3$ galaxies were taken from
Kriek et al.\ (2008a) and corrected to a {Kroupa} (2001) IMF.
The median mass of the $z\sim 2.3$ galaxies is $1.7
\times 10^{11}$\,M$_{\odot}$.
The median $r_e$ of SDSS red sequence galaxies
with masses $1.5-1.9 \times 10^{11}$\,M$_{\odot}$ is
5.0\,kpc, a factor of $\sim 6$ larger than the median size
of the $z\sim 2.3$ galaxies.

The combination of small sizes and high masses implies very
high densities.
The right panel of Fig.\ \ref{relations.plot} shows the
relation between stellar density and stellar mass,
with density defined as $\rho = 0.5 M / (\frac{4}{3}\pi r_e^3)$
(i.e., the mean stellar density within the effective radius, assuming
a constant stellar mass-to-light ($M/L$) ratio with radius).
The median density of the $z\sim 2.3$ galaxies
is $3\times 10^{10}$\,M$_{\odot}$\,kpc$^{-3}$ (with
a considerable rms scatter of 0.7 dex), a factor of
$\sim 180$ higher than the densities of local red sequence
galaxies of the same mass.

We note that it is difficult to determine the morphologies of the galaxies,
as they are so small. Nevertheless,
it is striking that several galaxies are quite elongated (see
Fig.\ \ref{stamps.plot}).
The most elongated galaxies are also the ones with the lowest
$n$ values (the correlation between $n$ and $b/a$ is
formally significant at the $>99$\,\% level\footnote{There
is no significant correlation between $r_e$ and $n$, or $r_e$ and
$b/a$.}), and
a possible interpretation is that the light of a subset of
the galaxies is dominated by very compact, massive disks (see
\S\,\ref{discussion.sec}).

\section{Discussion}
\label{discussion.sec}

We find that all ($100^{+0}_{-11}$\,\%) of the
quiescent, massive galaxies at $\langle z\rangle = 2.3$
spectroscopically identified by {Kriek} {et~al.} (2006) are extremely
compact, having a median effective radius of only 0.9\,kpc. This result
extends previous work at $z\sim 1.5$ ({Trujillo} {et~al.} 2007; Longhetti
et al.\ 2007; {Cimatti} {et~al.} 2008),
and confirms other studies at similar redshifts that were based on
photometric redshifts and images of poorer quality
({Zirm} {et~al.} 2007; {Toft} {et~al.} 2007).
Our study, together with the spectroscopy
in {Kriek} {et~al.} (2006) which demonstrates that the $H$-band light
comes from evolved stars, shows that the small measured sizes of
evolved high redshift galaxies are not caused by photometric redshift
errors, AGN, dusty starbursts, or measurement errors.

It is remarkable that all nine galaxies are so compact: even the largest
galaxy in the sample (HDFS1-1849) is significantly offset from the
relations of red galaxies in the nearby Universe (see Fig.\ 2).
We do not find any galaxy resembling a fully assembled elliptical or
S0 galaxy, which means that such objects make up less than $\sim 10$\,\% of the
population of  quiescent galaxies at $z\sim 2.3$.
This result effectively rules out simple monolithic models in which
early-type galaxies are assembled at the same time as their stars, and
is arguably the most direct evidence to date for an essentially
hierarchical assembly history for massive galaxies.

Our results are qualitatively consistent
with recent hydrodynamical
simulations in a $\Lambda$CDM Universe, which predict that
the stars in the central parts of massive elliptical galaxies formed
rapidly at $z\sim 5$ out of a concentrated gas distribution
(e.g., {Naab} {et~al.} 2007).
Observationally, it is tempting to link the compact
$z\sim 2.3$ galaxies to the rapidly rotating, dense
molecular gas reservoirs associated with some submm
galaxies\footnote{Note, however, that these submm galaxies
would have to be at significantly higher redshift
than the objects discussed in Tacconi et al.}
(Tacconi et al.\ 2008) and high redshift quasars
(e.g., {Riechers} {et~al.} 2006; {Narayanan} {et~al.} 2008). The fact that
several
of the galaxies in our sample (and the two galaxies
identified by {Stockton} {et~al.} 2008) appear to be disks provides
indirect support for this idea. 

Various mechanisms can play a role in bringing the $z\sim 2.3$
galaxies onto the scaling relations followed by today's red
galaxies. The total mass in galaxies on
the red sequence increases by a factor of $\sim 8$ from $z\sim 2.3$
to $z=0$ (Kriek et al.\ 2008b), and the galaxies that are
added after $z\sim 2.3$ may have preferentially larger sizes than
the galaxies that are already in place at that redshift. We note,
however, that
the galaxies described here are quite extreme even when compared to the
most compact 10\,\% of today's massive galaxies. Furthermore,
mass loss from stellar winds will decrease the stellar mass
(and increase the radius) from
$z=2.3$ to $z=0$.
Finally, accretion of satellites (e.g., {Naab} {et~al.} 2007)
and dry mergers (e.g., {van Dokkum} 2005)
may increase the sizes over time. 
Mergers will, to first order, move galaxies
roughly parallel to the size -- mass relation,
but the effects
on the sizes may be stronger in simulations with
realistic boundary conditions (see {Naab} {et~al.} 2007). Also,
the smallest galaxies are likely to merge with larger galaxies,
and even a merger between two small galaxies will (obviously)
reduce their number. Each of these mechanisms could plausibly
alter the size -- mass relation by a factor of 1.5--2, but not
a factor of $\sim 6$. This means
that some combination of effects
is required to bring the compact $z\sim 2.3$ galaxies to
the local relations --- or that we have not yet identified
the main mechanism.

We also cannot exclude the possibility that we are overestimating
the densities of these galaxies, as there are several systematic
uncertainties in the analysis. First, we may be underestimating
the sizes of the objects if their morphologies are complex.
In particular, an extended low surface brightness component could
be ``hiding'' in the noise, in which case the measured
effective radius would be larger if we had data of higher S/N.
There is no evidence for this from fitting a Sersic profile
to the stacked image
of the nine galaxies (see Fig.\ \ref{stamps.plot}), but we do note
that the galaxies are on average
$\sim 0.2$ mag brighter in ground-based
images (which include any extended emission) than in the
HST images.
A more subtle effect is an age-induced radial
gradient in the $M/L$ ratio of the galaxies, such that the central
parts are younger than the outer parts.
If this is the case the mass-weighted $r_e$
could be significantly larger than the luminosity-weighted $r_e$.
Such a spatial arrangement arises
in several different classes
of models for the formation of massive ellipticals
({Robertson} {et~al.} 2006; {Naab} {et~al.} 2007, P.\ Hopkins et al.,
in preparation). Lastly,
the IMF might be weighted toward high mass stars in these systems.
Several
studies have argued that the IMF in the progenitors of elliptical
galaxies may have been deficient in low mass stars, which would
impact their inferred masses and densities
(e.g., {Larson} 2005, van Dokkum 2008). The precise effect depends on the
assumed form of the IMF, and can be a factor of $\sim 2$
for plausible parameters (see van Dokkum 2008).

There are several ways to improve upon the results presented here.
Radial gradients in the $M/L$ ratio can, in principle, be measured from
high quality imaging in multiple passbands. Very deep imaging
can provide better constraints on the surface brightness
profiles at large radii. Ultimately, kinematics should provide
the most direct information on the nature of these compact
objects (see also Toft et al.\ 2007, Cimatti et al.\ 2008).
As shown by the broken lines in Fig.\ 2 (derived from
the relation between $M$, $\sigma$, and $r_e$ given in van
Dokkum \& Stanford 2003) the implied velocities are very high at
300 -- 500\,\kms. Finally, it will be interesting to build up
spectroscopic samples that are fainter in $K$, as this will allow us
to extend the
analysis to lower masses and to determine the
sizes of quiescent massive galaxies with higher $M/L$ ratios than the
objects studied here.

\acknowledgements{
We thank the referee, Ulrich Hopp, for his comments.
Support from NASA grant HST-GO-10808.01-A is gratefully acknowledged.
}


\end{document}